# Understanding electron correlation energy through density functional theory

by Teepanis Chachiyo[1,2,*] and Hathaithip Chachiyo[2]

[1] Department of Physics, Faculty of Science, Naresuan University, Phitsanulok 65000 Thailand
email: teepanisc@nu.ac.th

[2] Thailand Center of Excellence in Physics, Commission on Higher Education, 328 Si Ayutthaya Road, Bangkok 10400, Thailand    email: hathaithip.chachiyo@gmail.com

[*] Corresponding author

## Abstract

A curious behavior of electron correlation energy is explored. Namely, the correlation energy is the energy that tends to drive the system toward that of the uniform electron gas. As such, the energy assumes its maximum value when a gradient of density is zero. As the gradient increases, the energy is diminished by a gradient suppressing factor, designed to attenuate the energy from its maximum value similar to the shape of a bell curve. Based on this behavior, we constructed a very simple mathematical formula that predicted the correlation energy of atoms and molecules. Combined with our proposed exchange energy functional, we calculated the correlation energies, the total energies, and the ionization energies of test atoms and molecules; and despite the unique simplicities, the functionals' accuracies are in the top tier performance, competitive to the B3LYP, BLYP, PBE, TPSS, and M11. Therefore, we propose that, as guided by the simplicities and supported by the accuracies, the correlation energy is the energy that locally tends to drive the system toward the uniform electron gas.

# 1. INTRODUCTION

The term "correlation energy" was introduced by Wigner [1] in the 1930s who pointed out its significance in the field of solid state physics. In quantum mechanics, the method of computing the correlation energy is rooted on a perturbation theory [2]; and there are many methods which can calculate the energy very accurately such as the Coupled-Cluster theory [3], Configuration Interaction [2], and Moller-Plesset perturbation theory [4]. In contrast to a perturbation theory where the correlation energy involves a set of molecular orbitals; density functional theory [5] states that the energy depends explicitly on the density of electrons.

Over the past decades, much has been understood regarding the behavior of exchange and correlation energies in density functional theory, especially their asymptotic limits which were used to construct successful functionals such as Becke-88, and PBE. Since the original idea was conceived in the 1920s by Thomas and Fermi [6], progress has been made continuously in the field including the seminal work of Becke [7] in 1988 which brought the error of the exchange energy down to less than 1% (as compared to the exact Hartree-Fock exchange). The Becke-88 functional is also the key ingredient in constructing the B3 hybrid functional [8], the top 10 most cited paper of all time [9]. The equivalently notable success is the non-empirical PBE exchange and correlation functionals [10] whose parameters are derived from seven theoretical criteria. An even larger set of criteria, 17 conditions, is satisfied by the SCAN functional [11] which belongs to the state-of-the-art meta-GGA group. The meta-GGA functionals [12] use the electron density $\rho(\vec{r})$, its gradient $\vec{\nabla}\rho(\vec{r})$, and the positive orbital kinetic energy density $\sum_n \frac{1}{2} |\vec{\nabla}\psi_n|^2$ to evaluate the exchange and correlation energy. Continuous advances are also being made in the high temperature conditions [13].

However, in addition to the asymptotic limits, the behavior in the intermediate region should be as equally important. As such, two functionals obeying similar limits could give different predictions if they behave differently in the region connecting those limits. In this study, we explore alternative ideas that lead to simple and accurate correlation functional. Tests on atoms and molecules are presented. We further discuss how the functional is related to the conditions obeyed by the PBE functional.

To gain an initial insight into how electron correlation affects electron density, we first consider an exactly solvable model where there is no need to approximate the solution neither through a series of perturbations, nor a set of predefined basis functions. Shown in figure 1(a) is the "Hooke atom" [14], where the two electrons are bound to the origin by a harmonic potential $V(r) = \frac{1}{2}kr^2$. For $k = \frac{1}{4}$, the Schrödinger equation was solved exactly in 1990s by Kais et al. yielding the ground state total energy and the ground state wave function [15].

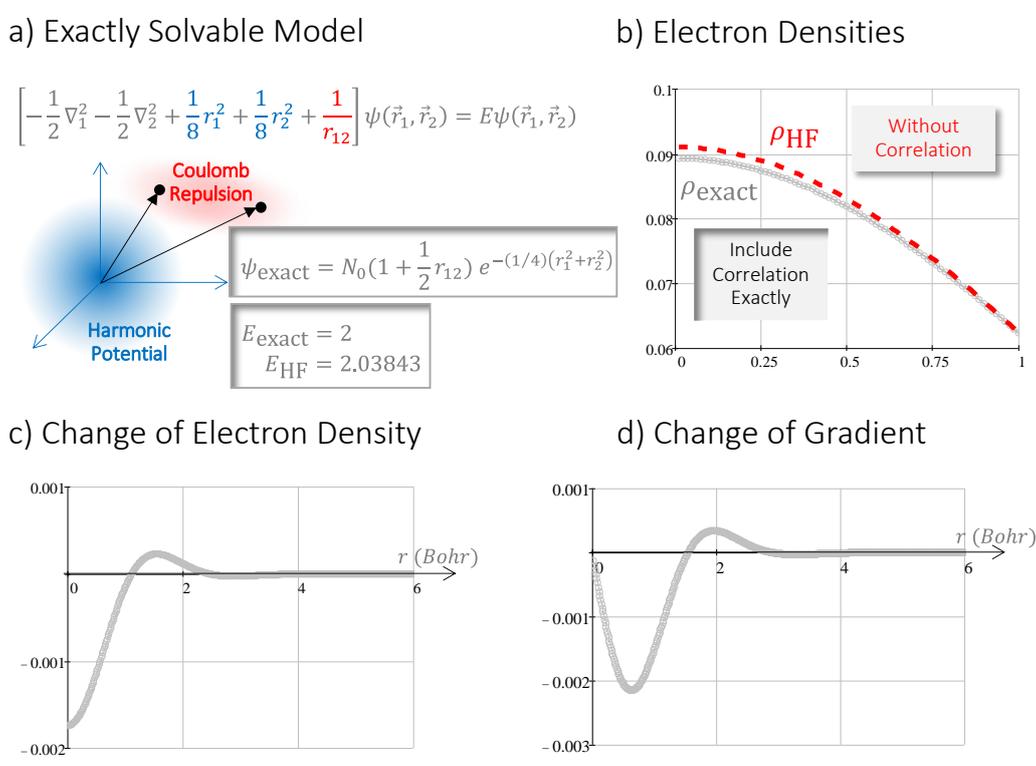

**Figure 1.** a) Hooke atom. b) Electron densities. The Hartree-Fock calculation is reproduced as described by Ref. [15]. c) Change of electron density due to correlation: $\rho_{\text{exact}} - \rho_{\text{HF}}$. d) Change of gradient due to correlation: $\left|\vec{\nabla}\rho\right|_{\text{exact}} - \left|\vec{\nabla}\rho\right|_{\text{HF}}$.

Figure 1(b) shows that the exact density changes less rapidly than the Hartree-Fock density. Since the Hartree-Fock method contains no correlation; it must be the effect of correlation in the exact density that drives the system closer to homogeneity. The change of gradient in figure 1(d) is also predominantly negative, meaning the overall gradient of the system is reduced due to electron correlation. In this specific example, the exactly solvable solution shows that electron correlation favors more homogeneous distribution of electrons.

Is it possible that the tendency to drive the system toward homogeneity is also a characteristic of electron correlation in general, applicable to all atoms and molecules? In this work, we entertain such hypothesis from which a mathematical correlation functional can be derived. The accuracy of the functional will then hint our understanding of the electron correlation energy, and how it affects the density of electrons in general.

## 2. METHODS

We hypothesize that the correlation energy is the energy that tends to drive the system toward the uniform electron gas. The hypothesis leads to a very simple expression as follows.

$$E_c = \int \rho \varepsilon_c (1 + t^2)^{h/\varepsilon_c} d^3r \qquad (1)$$

Here, $\rho$ is the total electron density; and $\varepsilon_c$ is the correlation energy when the density is perfectly uniform, which we use the Chachiyo's formula [16-18] $\varepsilon_c = a \ln(1 + \frac{b}{r_s} + \frac{b}{r_s^2})$. The term $(1 + t^2)^{h/\varepsilon_c}$ is the gradient suppressing factor, designed to attenuate the energy from its maximum value as a gradient parameter [19] $t = \left(\frac{\pi}{3}\right)^{1/6} \frac{\sqrt{a_0}}{4} \frac{|\vec{\nabla}\rho|}{\rho^{7/6}}$ increases. In addition, the constant $h = \frac{1}{2}(8.470 \times 10^{-3}) \times 16 \left(\frac{3}{\pi}\right)^{1/3} = 0.06672632$ Hartree is related to the behavior of the correlation energy when electron density varies very slowly.

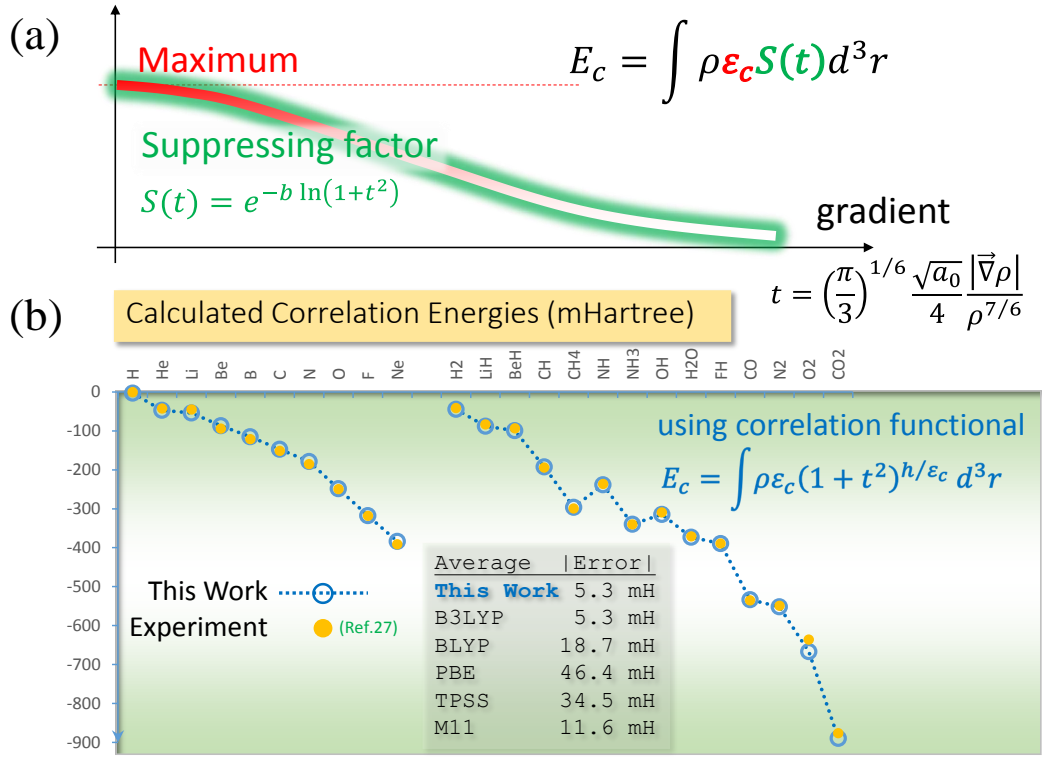

**Figure 2.** (a) Diagram showing how the strength of the correlation energy is suppressed as the gradient parameter $t$ increases. (b) Correlation energies predicted by the functional in this work. Basis set QZP was used throughout.

The derivation of Eq. (1) is simple but not rigorous. As shown in figure 2(a) if we conjecture that the correlation energy favors the uniformity of the electron density, then it must assume a maximum value when the gradient parameter $t = 0$. As $t$ increases, the energy is diminished which we initially accounted for it by using a "gradient suppressing factor" $S(t)$ in a form of Gaussian decay.

$$\text{Initial Form:} \quad S(t) = e^{-bt^2} \tag{2}$$

It was exactly the shape of a bell curve, monotonically decaying as $t$ increased. It was also consistent with the slowly varying density limit in which Ma and Brueckner [20] derived that

$$E_c \to \int \left[\rho\varepsilon_c + 8.470 \times 10^{-3} \frac{|\vec{\nabla}\rho|^2}{\rho^{4/3}}\right] d^3r \quad \text{Rydberg} \quad \text{as } |\vec{\nabla}\rho| \to 0 \tag{3}$$

As the gradient approached zero, the $e^{-bt^2} \to (1 - bt^2)$. Therefore, in this limit

$$\int \rho \varepsilon_c S(t) \, d^3r \to \int \rho \varepsilon_c [1 - bt^2] \, d^3r = \int \left[ \rho \varepsilon_c - \varepsilon_c b \left(\frac{\pi}{3}\right)^{1/3} \frac{1}{16} \frac{|\vec{\nabla}\rho|^2}{\rho^{4/3}} \right] d^3r, \quad (4)$$

where the factor $\left(\frac{\pi}{3}\right)^{1/3} \frac{1}{16}$ came from the definition of $t$. Demanding that Eq. (3) and Eq. (4) agreed, we had

$$b = -h/\varepsilon_c, \text{ and } h = \frac{1}{2}(8.470 \times 10^{-3}) \times 16 \left(\frac{3}{\pi}\right)^{1/3} = 0.06672632 \text{ Hartree} \quad (5)$$

However, the $t^2$ dependence in the initial form $S(t) = e^{-bt^2}$ was not appropriate when the gradient became larger. As pointed out by Perdew, Burke, and Ernzerhof when they constructed the highly successful non-empirical PBE [10] functional, the dependence ought to be proportional to $\ln t^2$ in order to cancel the logarithmic divergence of the uniform electron gas correlation energy.

In this work, we reconciled the two domains by using $\ln(1 + t^2)$ instead. That way, it converged to $t^2$ when $t$ was small, and approached $\ln t^2$ when $t$ was large. Therefore, we arrived at the

$$\text{Final Form:} \quad S(t) = e^{-b \ln(1+t^2)} = e^{\frac{h}{\varepsilon_c} \ln(1+t^2)}, \quad (6)$$

which promptly led to the correlation functional in Eq. (1).

In the presence of a spin polarization $\zeta = \frac{\rho_\alpha - \rho_\beta}{\rho}$, the uniform electron gas correlation energy $\varepsilon_c$ could be written as an interpolation between two extreme cases: the paramagnetic and the ferromagnetic.

$$\varepsilon_c(r_s, \zeta) = \varepsilon_c^0 + (\varepsilon_c^1 - \varepsilon_c^0)f(\zeta) \tag{7}$$

In the original paper [16], judging from the similar $\frac{1}{r_s}$ behavior in the low density limit, we had suggested the vonBarth-Hedin's $f_{\text{vBH}}(\zeta) = \frac{(1+\zeta)^{4/3}+(1-\zeta)^{4/3}-2}{2(2^{2/3}-1)}$ initially developed for the exchange energy [21] be used as the weighting function for the correlation energy as well.

In 1991 Wang and Perdew [22] had studied how the spin polarization affected the correlation energy $\varepsilon_c$ in the high density limit. They found that the vBH was not as accurate, and proposed a spin scaling factor $I_{\text{Wang-Perdew}} = g^3(\zeta)$ where $g(\zeta) = \frac{(1+\zeta)^{2/3}+(1-\zeta)^{2/3}}{2}$.

However, Wang and Perdew casted their formalism in terms of a scaling *factor*; but what we needed was a better *weighting function*. As such, we turned the scaling factor $g^3(\zeta)$ into a weighting function $f(\zeta)$ in the same way (but in reverse) they turned vBH weighting function into a factor ($I_{\text{vBH}} = 1 - \frac{1}{2}f_{\text{vBH}}(\zeta)$, see Eq. 43 in Wang and Perdew, 1991).

Therefore, in this work we propose using

$$f(\zeta) = 2(1 - g^3(\zeta)) \tag{8}$$

as a weighting function when taking into account of a spin polarization in a system.

To produce the results as shown in the following section one also needs an exchange energy contribution which we used our previously proposed exchange functional [23].

$$E_x = \int \rho \varepsilon_x \frac{3x^2 + \pi^2 \ln(x+1)}{(3x+\pi^2)\ln(x+1)} d^3r. \tag{9}$$

Here, $\varepsilon_x$ is the Dirac exchange energy [24] for uniform electron gas; and the parameter $x = \frac{|\vec{\nabla}\rho|}{\rho^{4/3}} \frac{2}{9}\left(\frac{\pi}{3}\right)^{1/3}$ is used to represent the gradient of the electron density.

Recently, the functional was also implemented in the LibXC [25] under the name "Chachiyo exchange". It was also used to compute electron densities in a newly proposed method for building initial guesses called SAP (Superposition of Atomic Potential) [26]. The study reported that, out of hundreds functionals available in LibXC, the Chachiyo exchange yielded the best initial guess wave functions on average.

## 3. RESULTS AND DISCUSSIONS

**Table 1. Total energies (Hartree)**

|      | Exact (Ref.27) | This Work Δ |
|------|---------------|-------------|
| H2   | -1.175        | -0.004      |
| LiH  | -8.070        | -0.004      |
| BeH  | -15.247       | -0.006      |
| CH   | -38.479       | +0.003      |
| CH4  | -40.516       | +0.003      |
| NH   | -55.223       | -0.001      |
| NH3  | -56.565       | +0.002      |
| OH   | -75.737       | -0.003      |
| H2O  | -76.438       | +0.001      |
| FH   | -100.459      | +0.002      |
| CO   | -113.326      | +0.008      |
| N2   | -109.542      | +0.009      |
| O2   | -150.327      | -0.019      |
| CO2  | -188.601      | -0.005      |
|            | This Work | 5.0 mH  |
|            | B3LYP   | 6.4 mH  |
| Average \|Δ\| | BLYP | 20.3 mH |
|            | PBE     | 55.4 mH |
|            | TPSS    | 40.1 mH |
|            | M11     | 9.6 mH  |

**Table 2. Ionization energies (eV)**

|     | Exp. (Ref.28) | This Work Δ |
|-----|---------------|-------------|
| H   | 13.60         | +0.09       |
| He  | 24.59         | +0.09       |
| Li  | 5.39          | +0.13       |
| Be  | 9.32          | -0.44       |
| B   | 8.30          | +0.32       |
| C   | 11.26         | +0.20       |
| N   | 14.53         | +0.10       |
| O   | 13.62         | +0.30       |
| F   | 17.42         | +0.08       |
| Ne  | 21.56         | -0.07       |
| Na  | 5.14          | +0.06       |
| Mg  | 7.65          | -0.23       |
| Al  | 5.99          | +0.08       |
| Si  | 8.15          | +0.00       |
| P   | 10.49         | -0.05       |
| S   | 10.36         | +0.01       |
| Cl  | 12.97         | -0.08       |
| Ar  | 15.76         | -0.14       |
|            | This Work | 0.14 |
|            | B3LYP     | 0.15 |
| Avg.\|Δ\|  | BLYP      | 0.18 |
|            | PBE       | 0.15 |
|            | TPSS      | 0.13 |
|            | M11       | 0.14 |

Figure 2(b), Table 1, and Table 2 illustrate the accuracy of the correlation functional in Eq. (1) via the calculated correlation energies, the total energies, and the ionization energies of atoms and molecules. The performance is comparable to that of B3LYP [8], BLYP [7,29], PBE [10], TPSS [12], and M11 [30]. Considering that the five functionals are among the most cited works of all time [9], the results validate the accuracy of the correlation energy functional in Eq. (1).

Owing to the accuracy of the energy functional in Eq. (1), we further discuss the implications of the hypothesis on the meaning of the correlation energy, both qualitatively and quantitatively.

Qualitatively, it is well known that the correlation energy plays a pivotal role in chemical bonding. This can easily be explained if one subscribes to the idea that the correlation energy favors uniform electron distribution. In the bonding region, the electron density hangs in balance between the two attractive nuclei; so the density becomes mostly uniform in this middle ground. Hence, the correlation energy approaches the maximum value and is expected to contribute appreciably in the bonding region.

Quantitatively, it has been noticed [31] that Local Density Approximation (LDA) theory [6] overestimates the correlation energy roughly by a factor of 2. The LDA is almost identical to the expression in Eq. (1) except that LDA does not have the gradient suppressing factor. In other words, the predicted correlation energy is always at its maximum value for LDA. Since the gradient suppressing factor $S(t)$ ranges from zero to one, a rough estimate would be a factor of 1/2, which would have brought down the LDA's overestimated value roughly to the right one.

We now discuss the results in figure 2(b) and Table 1, showing predictions from B3LYP, BLYP, PBE, M11, TPSS, and the functional in this work. Instead of the average errors, the differences are more pronounced if one focuses on the systematic errors in figure S1 and S2 in the supplementary material, where various functionals consistently overestimate (or underestimate) the correlation and the total energies. We attribute these differences to the intermediate region connecting the asymptotic limits, and how the functionals are constructed to satisfy those limits.

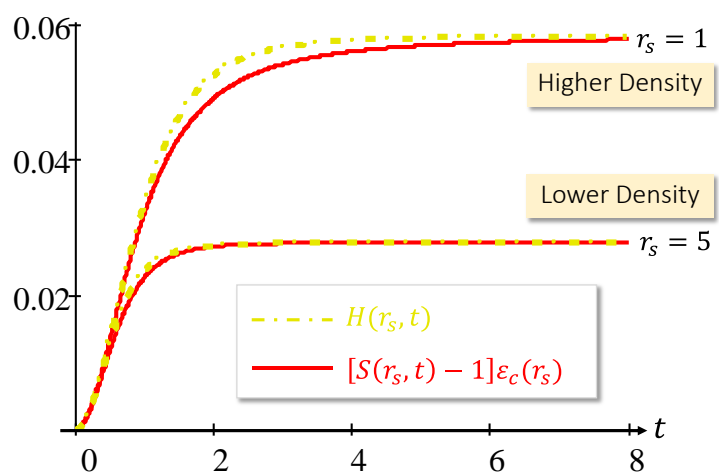

**Figure 3.** Comparison between the $H(r_s, t)$ from the PBE correlation and $[S(r_s, t) - 1]\varepsilon_c(r_s)$ for two different electron densities.

The correlation functional in this work follows the three conditions used in the PBE correlation. The difference is in the way we mathematically construct our functional. PBE uses the function $H(r_s, t)$ *to add* to the uniform electron gas correlation energy, while we propose a function $S(r_s, t)$ that *multiplies* to it instead. The mathematical forms are also different. Figure 3 shows how the two functionals respond to the gradient parameter $t$. Note how the two functionals obey the same asymptotic conditions, but behave differently in the intermediate region, especially when the electron density is higher. Therefore, we attribute the success of our correlation functional to its behavior in the intermediate region.

The tendency toward slowly varying electron gas, however, is not the only preferential behavior due to electron correlation. Because the strength of uniform electron gas correlation energy $\varepsilon_c(r_s)$ increases as a function of electron density; the system also prefers higher density. Since the total number of electrons is conserved via normalization, accumulating higher density in one region implies lower density in the adjacent areas, promoting higher gradient of electron density. Therefore, the tendency toward higher density may sometimes compete with the tendency toward homogeneity.

Most recently, it is shown [32] that for the large-Z limit of neutral atoms, electron correlation becomes local and slowly-varying. Further study is needed to address how the functional in this work performs in such limit, along with other important cases such as van der Waals interaction, hydrogen bonding, or bulk crystal.

## 4. CONCLUSIONS

Finally, we would like to emphasize that the simple behavior of the correlation energy in this work is not an oversimplification, but rather an accurate description of nature, as evident from the results in figure 2(b), Table 1, and 2. At first glance, it may seem very surprising that one could bypass all the complexities of the perturbation theory, and depend directly on the electron density in order to compute the correlation energy. The fact that this can be achieved as illustrated in Eq. (1) is a result of the fundamental premise of DFT, that the total energy can always be written as a functional of electron density. The true surprise nobody in the 90 year history of quantum mechanics would have suspected, however, is that the functional turns out to be very simple both in the way it can be calculated and interpreted. We propose that the correlation energy is the energy that locally tends to drive the system toward the uniform electron gas.

**Supplementary Material**

See supplementary material for additional methods for spin polarized cases, computational details, supplementary figures, and computer source codes.

**Acknowledgements**

We would like to thank Prof. Peter M. W. Gill for providing comments on the manuscript, and Prof. John P. Perdew for pointing out the competing effects of GGA correlation functional (private communication). Support by Thailand Center of Excellence in Physics grant No. ThEP-60-PET-NU9 is gratefully acknowledged.

# Supplementary Material

"Understanding electron correlation energy through density functional theory"

by Teepanis Chachiyo[1,2,*] and Hathaithip Chachiyo[2]

[1] Department of Physics, Faculty of Science, Naresuan University, Phitsanulok 65000 Thailand

[2] Thailand Center of Excellence in Physics, Commission on Higher Education, 328 Si Ayutthaya Road, Bangkok 10400, Thailand

[*] Correspondence to: <teepanisc@nu.ac.th>

## Supplementary Method

In this work, we implemented the proposed exchange and correlation functional given in Eq. (9) and Eq. (1) respectively in Siam Quantum[1] software package. The effect of spin polarization $\zeta = \frac{\rho_\alpha - \rho_\beta}{\rho}$ was taken into account via the uniform electron gas correlation energy $\varepsilon_c(r_s, \zeta) = \varepsilon_c^0 + (\varepsilon_c^1 - \varepsilon_c^0)f(\zeta)$. However, instead of using the vonBarth-Hedin weighting function $f(\zeta)$ as previously suggested[2], in this work we developed a new weighting function $f(\zeta) = 2(1 - g^3(\zeta))$ based on the spin-scaling factor[3] $g(\zeta) = \frac{(1+\zeta)^{2/3} + (1-\zeta)^{2/3}}{2}$ in the high density limit. For the exchange energy, the effect of spin polarization was taken into account by computing each spin separately, $E_x[\rho_\alpha, \rho_\beta] = \frac{1}{2}E_x[2\rho_\alpha] + \frac{1}{2}E_x[2\rho_\beta]$. The basis set QZP[4,5] (without G, H orbitals) was used throughout. In Siam Quantum, a basis set was expanded into 6D/10F Cartesian functions; and a numerical grid (75 radial, and 302 Labedev angular) was used in DFT calculations.

The reported B3LYP, BLYP, PBE, TPSS, and M11 energies were calculated using GAMESS.[6]

In Fig. 1(b), the correlation energies were computed by $E_{total}^{(DFT)} - E^{(HF)}$; where $E_{total}^{(DFT)}$ was the total energies from the respective DFT methods; and $E^{(HF)}$ was the Hartree-Fock energy calculated under the same basis set. The molecular geometries were taken from the G2 archive[7] which had been optimized by the MP2/6-31G* level of theory.

## Energies Calculation

Unless explicitly stated, atomic units were used throughout this work. Knowing the spin up $\rho_\alpha(\vec{r})$ and the spin down $\rho_\beta(\vec{r})$ electron density, we calculated the correlation energy using the expression in Eq. (1):

$$E_c = \int \rho \varepsilon_c (1 + t^2)^{h/\varepsilon_c} d^3 r \tag{S1}$$

Here, $\rho = \rho_\alpha + \rho_\beta$ was the total density; $t = \left(\frac{\pi}{3}\right)^{1/6} \frac{1}{4} \frac{|\vec{\nabla}\rho|}{\rho^{7/6}}$ and $h = 0.06672632$ Hartree. The effect of spin polarization $\zeta = \frac{\rho_\alpha - \rho_\beta}{\rho}$ was taken into account via the uniform electron gas correlation energy.

$$\varepsilon_c(r_s, \zeta) = \varepsilon_c^0 + (\varepsilon_c^1 - \varepsilon_c^0) f(\zeta) \tag{S2}$$

$\varepsilon_c^0$ and $\varepsilon_c^1$ were the Chachiyo's formula[2] for the paramagnetic and ferromagnetic case respectively.

$$\varepsilon_c^0 = a \ln\left(1 + \frac{b}{r_s} + \frac{b}{r_s^2}\right) \qquad a = \frac{\ln 2 - 1}{2\pi^2}; \quad b = 20.4562557 \tag{S3}$$

$$\varepsilon_c^1 = a \ln\left(1 + \frac{b}{r_s} + \frac{b}{r_s^2}\right) \qquad a = \frac{\ln 2 - 1}{4\pi^2}; \quad b = 27.4203609 \tag{S4}$$

However, instead of using the vonBarth-Hedin weighting function $f(\zeta)$ as previously suggested[2], in this work we developed a new weighting function

$$f(\zeta) = 2(1 - g^3(\zeta)) \tag{S5}$$

based on the spin-scaling factor[3] $g(\zeta) = \frac{(1+\zeta)^{2/3} + (1-\zeta)^{2/3}}{2}$ in the high-density limit.

For the exchange energy in this work, we used the Chachiyo exchange,

$$E_x[\rho] = \int \rho \varepsilon_x \frac{3x^2 + \pi^2 \ln(x+1)}{(3x + \pi^2)\ln(x+1)} d^3r. \tag{S6}$$

Here, $\varepsilon_x[\rho] = -\frac{3}{4}\left(\frac{3}{\pi}\rho\right)^{1/3}$ was the Dirac exchange energy for uniform electron gas; and $x[\rho] = \frac{|\vec{\nabla}\rho|}{\rho^{4/3}} \frac{2}{9}\left(\frac{\pi}{3}\right)^{1/3}$. The effect of spin polarization was taken into account by computing each spin separately as follows.

$$E_x[\rho_\alpha, \rho_\beta] = \frac{1}{2}E_x[2\rho_\alpha] + \frac{1}{2}E_x[2\rho_\beta]; \tag{S7}$$

The integration for the exchange and correlation energy were computed numerically using quadrature method with 75 radial points and 302 Labedev angular points per atom.

$$\int f(\vec{r}) d^3r \cong \sum_n w_n f(\vec{r}_n) \tag{S8}$$

## Determination of Electron Densities

To compute the spin up $\rho_\alpha(\vec{r})$ and the spin down $\rho_\beta(\vec{r})$ electron density, we solved the Kohn-Sham equation[8] for both spin types, namely,

$$\left[-\frac{1}{2}\nabla^2 + \hat{v}_{ext} + \hat{v}_J + \hat{v}_x^{(\alpha)} + \hat{v}_c^{(\alpha)}\right]\phi_n^{(\alpha)} = \epsilon_n^{(\alpha)}\phi_n^{(\alpha)} \tag{S9}$$

and

$$\left[-\frac{1}{2}\nabla^2 + \hat{v}_{ext} + \hat{v}_J + \hat{v}_x^{(\beta)} + \hat{v}_c^{(\beta)}\right]\phi_n^{(\beta)} = \epsilon_n^{(\beta)}\phi_n^{(\beta)} \tag{S10}$$

The two eigen equations above yielded two sets of molecular orbitals $\{\phi_n^{(\alpha)}\}$ and, $\{\phi_n^{(\beta)}\}$ in which we used to compute the electron densities.

$$\rho_\alpha(\vec{r}) = \Sigma\left|\phi_n^{(\alpha)}\right|^2 \text{ and } \rho_\beta(\vec{r}) = \Sigma\left|\phi_n^{(\beta)}\right|^2 \tag{S11}$$

To solved the Eq. (S9) and Eq. (S10), we first derived the explicit expressions for the exchange potential $\hat{v}_x$ and correlation potential $\hat{v}_c$ consistent with the exchange and correlation energy functional in this work. For the exchange,

$$\hat{v}_x^{(\alpha)} = \hat{v}_x[2\rho_\alpha] \quad \text{and} \quad \hat{v}_x^{(\beta)} = \hat{v}_x[2\rho_\beta]; \tag{S12}$$

where the functional

$$\hat{v}_x[\rho] = \frac{4}{3}\varepsilon_x(F - xF') - \vec{\nabla}\cdot\vec{\Gamma} \tag{S13}$$

with the auxiliary terms: $F[\rho] = \frac{3x^2 + \pi^2 \ln(x+1)}{(3x+\pi^2)\ln(x+1)}$, $F'[\rho] = \frac{dF}{dx}$, $x[\rho] = \frac{|\vec{\nabla}\rho|}{\rho^{4/3}} \frac{2}{9} \left(\frac{\pi}{3}\right)^{1/3}$, and $\vec{\Gamma}[\rho] = -\frac{F'}{6|\vec{\nabla}\rho|}\vec{\nabla}\rho$.

For the correlation,

$$\hat{v}_c^\alpha = S\left[v_{c0}^\alpha - \frac{7}{3}\frac{ht^2}{1+t^2} - \left(\frac{v_{c0}^\alpha}{\varepsilon_c} - 1\right) h \ln(1+t^2)\right] - \vec{\nabla} \cdot \vec{\Gamma} \tag{S14}$$

$$\hat{v}_c^\beta = S\left[v_{c0}^\beta - \frac{7}{3}\frac{ht^2}{1+t^2} - \left(\frac{v_{c0}^\beta}{\varepsilon_c} - 1\right) h \ln(1+t^2)\right] - \vec{\nabla} \cdot \vec{\Gamma} \tag{S15}$$

Here, $v_{c0}^\alpha$ and $v_{c0}^\beta$ were the correlation potentials for the uniform electron gas for the spin-up and the spin-down respectively. The terms $S = (1+t^2)^{h/\varepsilon_c}$, $t = \left(\frac{\pi}{3}\right)^{1/6}\frac{1}{4}\frac{|\vec{\nabla}\rho|}{\rho^{7/6}}$, and $\vec{\Gamma} = 2S\frac{ht^2}{1+t^2}\frac{\rho}{|\vec{\nabla}\rho|^2}\vec{\nabla}\rho$ were evaluated using the total electron density. $\varepsilon_c$ was given in Eq. (S2).

Finally, the correlation potentials of the uniform electron gas were as follow.

$$v_{c0}^\alpha = \varepsilon_c - \frac{r_s}{3}\left[\frac{\partial \varepsilon_c^0}{\partial r_s} + \left(\frac{\partial \varepsilon_c^1}{\partial r_s} - \frac{\partial \varepsilon_c^0}{\partial r_s}\right) f(\zeta)\right] + 2(\varepsilon_c^1 - \varepsilon_c^0)\frac{\partial f}{\partial \zeta}\frac{\rho_\beta}{\rho} \tag{S16}$$

$$v_{c0}^\beta = \varepsilon_c - \frac{r_s}{3}\left[\frac{\partial \varepsilon_c^0}{\partial r_s} + \left(\frac{\partial \varepsilon_c^1}{\partial r_s} - \frac{\partial \varepsilon_c^0}{\partial r_s}\right) f(\zeta)\right] - 2(\varepsilon_c^1 - \varepsilon_c^0)\frac{\partial f}{\partial \zeta}\frac{\rho_\alpha}{\rho} \tag{S17}$$

$f(\zeta)$ was given in Eq. (S5); and $\frac{\partial f}{\partial \zeta} = -2g^2(\zeta)\left[(1+\zeta)^{-1/3} - (1-\zeta)^{-1/3}\right]$.

Having completely defined the potential, we proceeded to solve the Kohn-Sham equation in Eq. (S9) and Eq. (S10) via the basis function expansion method. Many sets of basis functions were available publicly. We downloaded it from the EMSL Basis Set Exchange database.[5]

# Supplementary Figures

These are four supplementary figures showing the errors of the calculated correlation, total, and ionization energies as presented in Fig. 1(b), Table I, and II respectively; and additionally atomization energies.

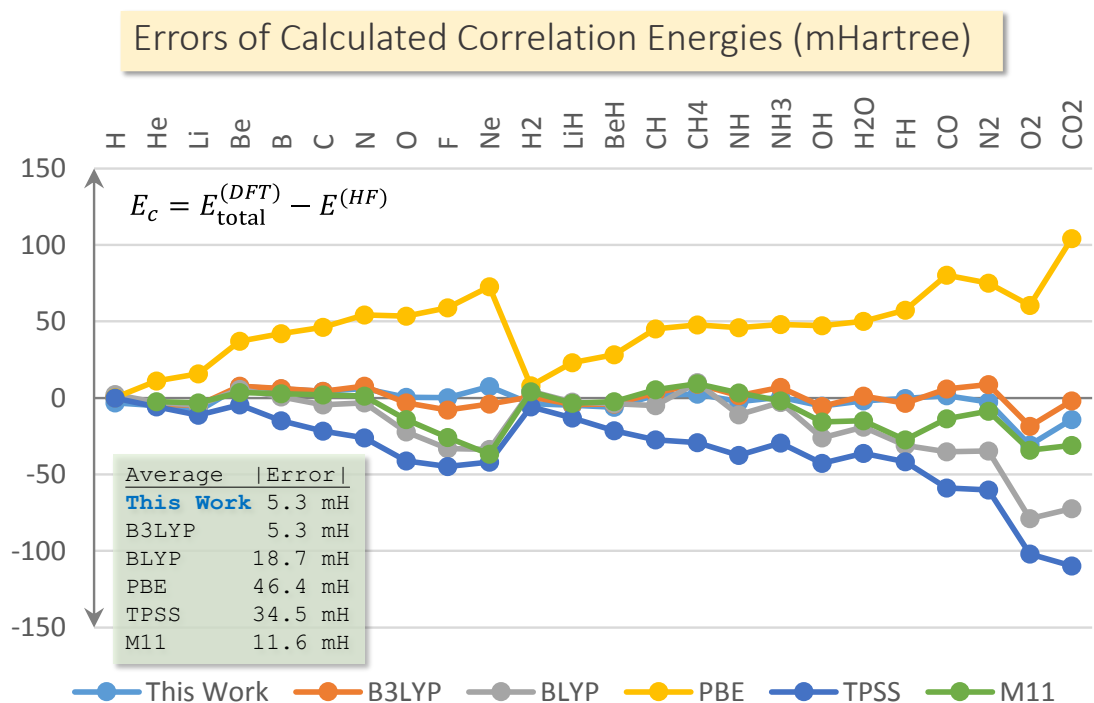

**Figure S1.** Errors of calculated correlation energies (mHartree).

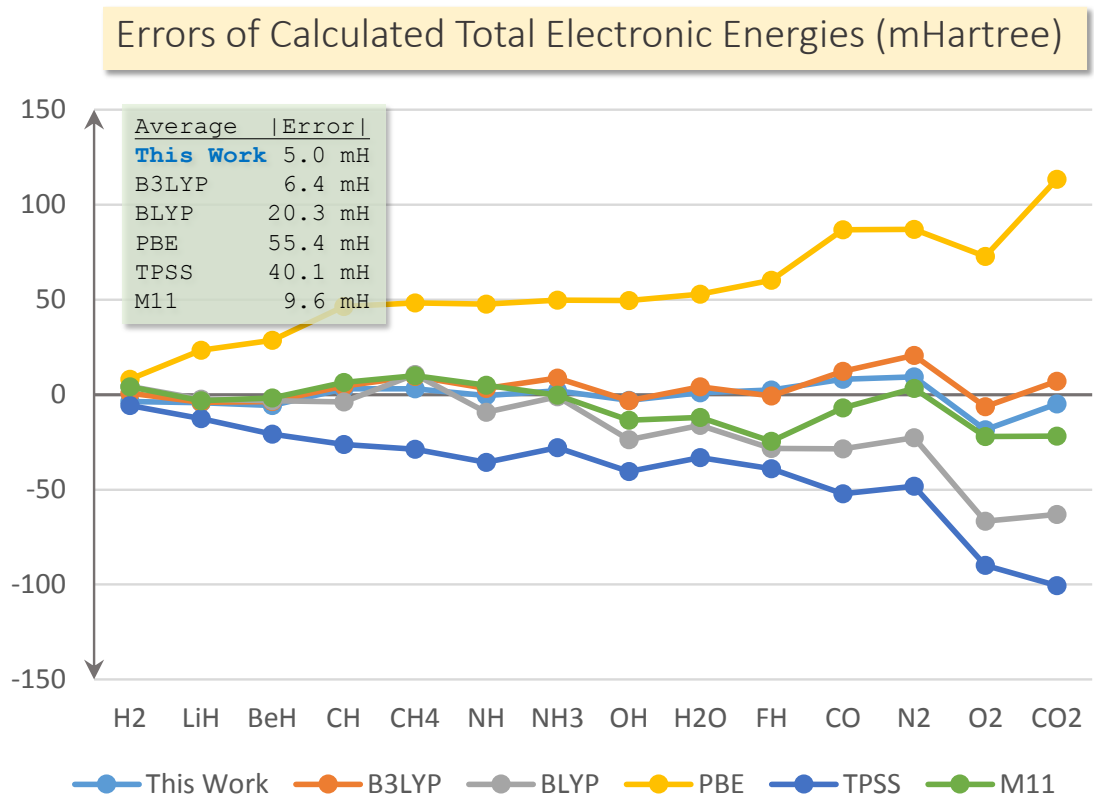

**Figure S2.** Errors of calculated total energies (mHartree).

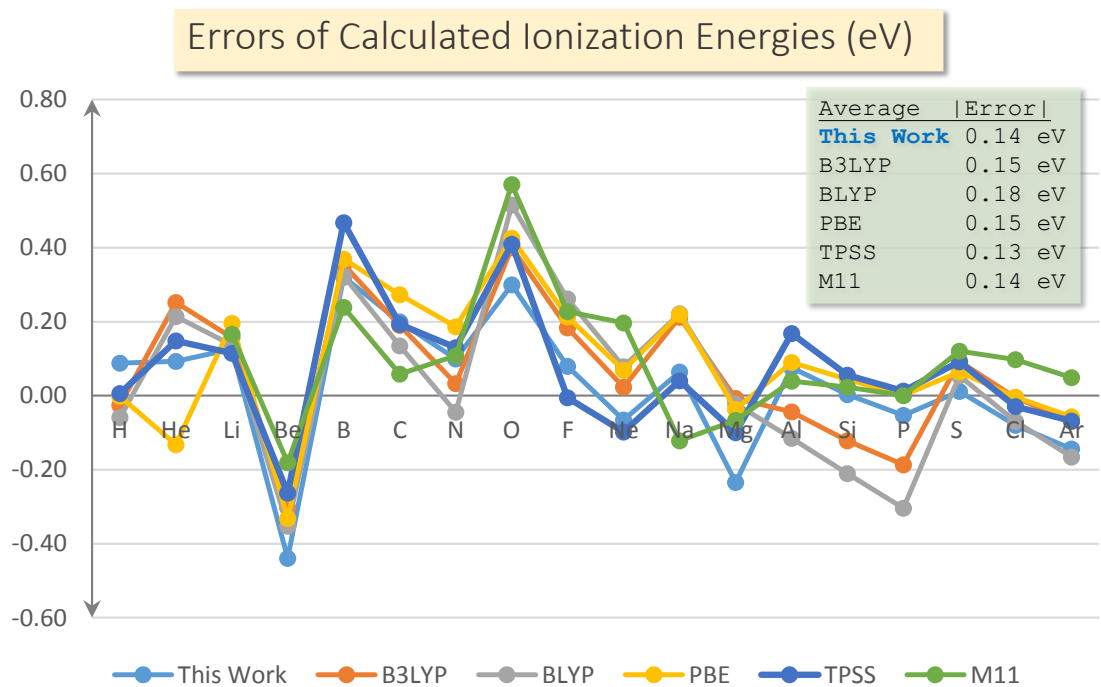

**Figure S3.** Errors of calculated ionization energies (eV).

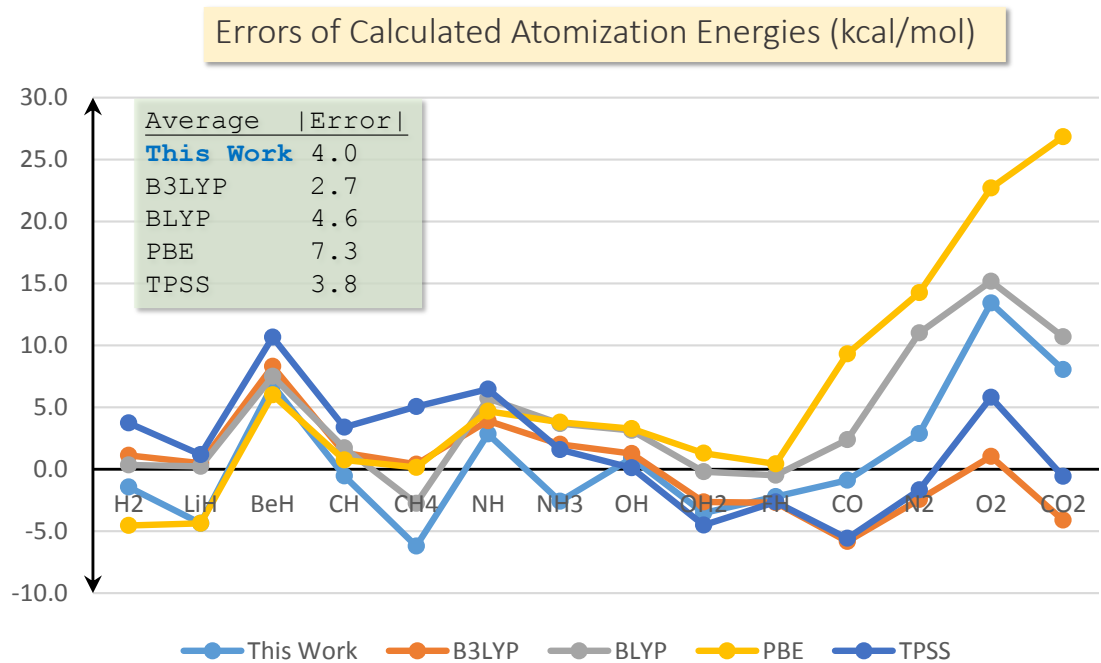

**Figure S4**. Errors of calculated atomization energies (kcal/mol).

Zero point energies and relativistic corrections have been subtracted from the experimental atomization values[9]. The calculated values are obtained by subtracting the total energies of constituent atoms from the total energy of the molecule.

# Geometries and energies of test molecules

The geometries were downloaded from http://www.cse.anl.gov, which was the archive for the MP/6-31G* G2 set. "Q" and "M" stand for molecular charge and spin multiplicity.

```
H2     Q=0    M=1
H     .000000     .000000      .368583
H     .000000     .000000     -.368583

LiH    Q=0    M=1
Li    .000000     .000000      .410000
H     .000000     .000000    -1.230000

BeH    Q=0    M=2
Be    .000000     .000000      .269654
H     .000000     .000000    -1.078616

CH     Q=0    M=2
C     .000000     .000000      .160074
H     .000000     .000000     -.960446

CH4    Q=0    M=1
C     .000000     .000000      .000000
H     .629118     .629118      .629118
H    -.629118    -.629118      .629118
H     .629118    -.629118     -.629118
H    -.629118     .629118     -.629118

NH     Q=0    M=3
N     .000000     .000000      .129929
H     .000000     .000000     -.909501
```

```
NH3    Q=0    M=1
N     .000000       .000000       .116489
H     .000000       .939731      -.271808
H     .813831      -.469865      -.271808
H    -.813831      -.469865      -.271808

OH     Q=0    M=2
O     .000000       .000000       .108786
H     .000000       .000000      -.870284

H2O    Q=0    M=1
O     .000000       .000000       .119262
H     .000000       .763239      -.477047
H     .000000      -.763239      -.477047

FH     Q=0    M=1
F     .000000       .000000       .093389
H     .000000       .000000      -.840502

CO     Q=0    M=1
O     .000000       .000000       .493003
C     .000000       .000000      -.657337

N2     Q=0    M=1
N     .000000       .000000       .564990
N     .000000       .000000      -.564990

O2     Q=0    M=3
O     .000000       .000000       .622978
O     .000000       .000000      -.622978

CO2    Q=0    M=1
C     .000000       .000000       .000000
O     .000000       .000000      1.178658
O     .000000       .000000     -1.178658
```

| | SQ/QZP (without G,H) | | |
|---|---|---|---|
| | HF | This Work | SVWN |
| H | -0.499884 | -0.502981 | -0.478316 |
| He | -2.861126 | -2.908144 | -2.833976 |
| Li | -7.432678 | -7.486382 | -7.343862 |
| Be | -14.572686 | -14.659240 | -14.446849 |
| B | -24.532690 | -24.647719 | -24.355486 |
| C | -37.693074 | -37.839803 | -37.469472 |
| N | -54.403539 | -54.582279 | -54.135517 |
| O | -74.817358 | -75.065859 | -74.529332 |
| F | -99.414207 | -99.731922 | -99.112258 |
| Ne | -128.544478 | -128.928094 | -128.230004 |
| H2 | -1.133533 | -1.178107 | -1.137201 |
| LiH | -7.987062 | -8.074712 | -7.919279 |
| BeH | -15.153339 | -15.252460 | -15.021810 |
| CH | -38.283553 | -38.475704 | -38.095028 |
| CH4 | -40.216235 | -40.512643 | -40.121074 |
| NH | -54.984976 | -55.223284 | -54.765674 |
| NH3 | -56.222959 | -56.562602 | -56.108173 |
| OH | -75.425758 | -75.740069 | -75.205327 |
| H2O | -76.064363 | -76.437314 | -75.910440 |
| FH | -100.067383 | -100.456808 | -99.848739 |
| CO | -112.784588 | -113.318063 | -112.474244 |
| N2 | -108.981226 | -109.532829 | -108.693714 |
| O2 | -149.678531 | -150.345215 | -149.334962 |
| CO2 | -187.715962 | -188.605964 | -187.281680 |

| | GAMESS/QZP (witout G,H) | | | | | | |
|---|---|---|---|---|---|---|---|
| | HF | B3LYP | BLYP | PBE | TPSS | M11 | SVWN |
| H | -0.499884 | -0.498785 | -0.497568 | -0.499659 | -0.499957 | NaN Error | -0.478316 |
| He | -2.861126 | -2.907349 | -2.906154 | -2.892063 | -2.908884 | -2.905466 | -2.833976 |
| Li | -7.432675 | -7.482481 | -7.482513 | -7.461989 | -7.488881 | -7.480765 | -7.343858 |
| Be | -14.572680 | -14.658991 | -14.661040 | -14.629506 | -14.671266 | -14.663145 | -14.446848 |
| B | -24.532693 | -24.647426 | -24.653070 | -24.611593 | -24.668543 | -24.650981 | -24.355486 |
| C | -37.693079 | -37.839600 | -37.848490 | -37.797796 | -37.865812 | -37.842124 | -37.469473 |
| N | -54.403541 | -54.580822 | -54.591757 | -54.534333 | -54.614632 | -54.587182 | -54.135520 |
| O | -74.817362 | -75.069635 | -75.088548 | -75.012813 | -75.107611 | -75.080400 | -74.529332 |
| F | -99.414209 | -99.739969 | -99.765307 | -99.673195 | -99.777006 | -99.758036 | -99.112260 |
| Ne | -128.544479 | -128.939336 | -128.969186 | -128.862643 | -128.977421 | -128.972134 | -128.230003 |
| H2 | -1.133533 | -1.173783 | -1.170106 | -1.166488 | -1.180290 | -1.170366 | -1.137201 |
| LiH | -7.987061 | -8.074361 | -8.072756 | -8.046998 | -8.083046 | -8.073532 | -7.919277 |
| BeH | -15.153340 | -15.250402 | -15.249957 | -15.218127 | -15.267601 | -15.248633 | -15.021812 |
| CH | -38.283555 | -38.474319 | -38.482612 | -38.432462 | -38.504977 | -38.472283 | -38.095027 |
| CH4 | -40.216240 | -40.506192 | -40.505192 | -40.467474 | -40.544489 | -40.505838 | -40.121177 |
| NH | -54.984978 | -55.219349 | -55.231938 | -55.174944 | -55.258382 | -55.217719 | -54.765675 |
| NH3 | -56.222964 | -56.555880 | -56.565826 | -56.514877 | -56.592525 | -56.564875 | -56.108166 |
| OH | -75.425759 | -75.740228 | -75.760897 | -75.687498 | -75.777551 | -75.750503 | -75.205326 |
| H2O | -76.064364 | -76.434098 | -76.454468 | -76.385297 | -76.471445 | -76.450347 | -75.910438 |
| FH | -100.067382 | -100.459883 | -100.487487 | -100.398932 | -100.498168 | -100.483767 | -99.848738 |
| CO | -112.784589 | -113.313733 | -113.354632 | -113.239238 | -113.378373 | -113.333177 | -112.474247 |
| N2 | -108.981227 | -109.521466 | -109.564775 | -109.455069 | -109.590321 | -109.538761 | -108.693713 |
| O2 | -149.678532 | -150.333015 | -150.393353 | -150.253887 | -150.416568 | -150.348653 | -149.334964 |
| CO2 | -187.715963 | -188.593948 | -188.664215 | -188.487783 | -188.701723 | -188.622999 | -187.281680 |

# Computer subroutines in C-language

These are 3 subroutines for computing the exchange and correlation energy as implemented in the Siam Quantum program.

Below is the subroutine for computing the correlation energy for non-uniform electron density. It calls a function "getDFT_cChachiyo" which evaluates the uniform electron gas correlation energy.

```c
// T.Chachiyo and H.Chachiyo (2019) "Understanding electron correlation energy
// through density functional theory"
//
void getDFT_cGGA_Chachiyo(
double rhoa,     // spin up   electron density
double rhob,     // spin down electron density
double rhog,     // total gradient
double *Ec,      // returned (incremental) correlation energy
double *dEdrhoa, // returned (incremental) spin up   potential
double *dEdrhob, // returned (incremental) spin down potential
double *G){      // returned (incremental) coef. of total density gradient

        double rho  = rhoa  + rhob;
        if(rho > RHO_CUTOFF){

                // From the Chachiyo correlation functional
                double va=0., vb=0., e_unif=0.;
                getDFT_cChachiyo(rhoa, rhob, &e_unif, &va, &vb);
                e_unif = e_unif/rho;

                double t = rhog/pow(rho,7.0/6) * pow(M_PI/3.0, 1.0/6)/4.0;
                double h = 0.06672632;
                double S = pow(1.0+t*t,h/e_unif);

                *dEdrhoa += S*(va-7.0/3*h*t*t/(1.0+t*t)-h*log(1.0+t*t)*(va/e_unif-1.0));
                *dEdrhob += S*(vb-7.0/3*h*t*t/(1.0+t*t)-h*log(1.0+t*t)*(vb/e_unif-1.0));

                if(rhog*rhog > RHO_CUTOFF)
                        *G += 2.0*h*S*t*t/(1.0+t*t)*rho/rhog/rhog;

                *Ec += rho * S * e_unif;
        }
}
```

Below is the function that evaluates the uniform electron gas correlation energy.

```c
void getDFT_cChachiyo(
//
// Teepanis Chachiyo (2016). "Simple and accurate uniform electron gas
// correlation energy for the full range of densities". J. Chem. Phys.
// 145 (2): 021101.
//
// T.Chachiyo and H.Chachiyo (2019) "Understanding electron correlation energy
// through density functional theory"
//
double rhoa,      // spin up   electron density
double rhob,      // spin down electron density
double *Ec,       // returned (incremental)  correlation energy
double *dEdrhoa,  // returned (incremental) spin up   potential
double *dEdrhob){// returned (incremental) spin down potential

	double rho=rhoa+rhob;
	if(rho>RHO_CUTOFF){

		double z = (rhoa-rhob)/rho;
		double g = (pow(1.+z,2./3)+pow(1.-z,2./3))/2.0;
		double f = 2.0*(1.0-g*g*g);
		double dfdz;
		if(fabs(1.0-z*z)>RHO_CUTOFF) dfdz = -2.0*g*g*(pow(1.+z,-1./3)-pow(1.-z,-1.0/3));
		else                         dfdz =  0.0;

#define Chachiyo_formula(a,b,rs) a*log(1.+b/rs+b/rs/rs)
#define decdrs(a,b,rs)           a/(1. + b/rs + b/rs/rs)*b*(-1./rs/rs-2./rs/rs/rs);
		double rs = pow(3./4/M_PI/rho,1./3);
		double a  = (log(2.)-1.)/M_PI/M_PI/2.;
		double b  = 20.4562557;
		double e0 = Chachiyo_formula(a,b,rs);
		double de0drs = decdrs(a,b,rs);
		       a  = (log(2.)-1.)/M_PI/M_PI/4.;
		       b  = 27.4203609;
		double e1 = Chachiyo_formula(a,b,rs);
		double de1drs = decdrs(a,b,rs)
		double e  = e0 + (e1-e0)*f;

		*dEdrhoa += e - rs/3.0*(de0drs + (de1drs-de0drs)*f) + (e1-e0)*dfdz*2.0*rhob/rho;
		*dEdrhob += e - rs/3.0*(de0drs + (de1drs-de0drs)*f) - (e1-e0)*dfdz*2.0*rhoa/rho;

		*Ec      += e * rho;
	}

}
```

S16Below is the code for computing the exchange functional in Eq. (9).

```c
// T.Chachiyo and H.Chachiyo (2019) "Understanding electron correlation energy
// through density functional theory"
//
void getDFT_xGGA_Chachiyo(
double rhoa,     // spin up   electron density
double rhob,     // spin down electron density
double rhoag,    // spin up   density gradient
double rhobg,    // spin down density gradient
double *Ex,      // returned (incremental) exchange energy
double *dEdrhoa, // returned (incremental) spin up   potential
double *dEdrhob, // returned (incremental) spin down potential
double *Ga,      // returned (incremental) coef. of alpha density gradient
double *Gb){     // returned (incremental) coef. of beta  density gradient

#define PI2  M_PI*M_PI
#define eDirac(rho) -3./4*pow(3./M_PI*rho,1./3)
#define GET_F    F    = (3.0*x*x+PI2*log(x+1.0))/(3.0*x+PI2)/log(x+1.0)
#define GET_dFdx dFdx = ( 6.0*x*(x+1.0) + PI2                           \
                        - F*(3.0*x + PI2 + 3.0*(x+1.0)*log(x+1.0)) )\
                       /(3.0*x+PI2)/(x+1.0)/log(x+1.0)

        double e,x,F,dFdx;

        // spin up contribution
        if(rhoa > RHO_CUTOFF){
                e = eDirac(2.0*rhoa);
                x = (2.0*rhoag)/pow(2.0*rhoa,4./3) * 2.0/9*pow(M_PI/3.,1./3);
                if(x<RHO_CUTOFF){ F = 1.0; dFdx = 0.0; } else{ GET_F; GET_dFdx; }
                *Ex      += rhoa  * e * F;
                *dEdrhoa += 4.0/3 * e * (F-x*dFdx);
                if(rhoag > RHO_CUTOFF) *Ga += -dFdx/rhoag/6.0;
        }

        // spin down contribution
        if(rhob > RHO_CUTOFF){
                e = eDirac(2.0*rhob);
                x = (2.0*rhobg)/pow(2.0*rhob,4./3) * 2.0/9*pow(M_PI/3.,1./3);
                if(x<RHO_CUTOFF){ F = 1.0; dFdx = 0.0; } else{ GET_F; GET_dFdx; }
                *Ex      += rhob  * e * F;
                *dEdrhob += 4.0/3 * e * (F-x*dFdx);
                if(rhobg > RHO_CUTOFF) *Gb += -dFdx/rhobg/6.0;
        }
}
```

# Hartree-Fock solution of Hooke atom

The solution has already been studied by Kais et al.[10] To reproduce their results and plot the graphs in figure 1, we followed the equations and the definitions in Eq. (9), (10), (11), and (12) in their paper. While the authors used the program CADPACS, the Cambridge Analytic Derivatives Package, to numerically solve the Eq. (9), we used a simple finite difference method. The results should not be significantly different. Our value for the Hartree-Fock energy was 2.0384 which agreed well with their value, 2.0393.

With 800 radial grid points, our subroutine was sufficient to reproduce the plot of FIG. 1 (upper graph) in their paper.

We also tested the subroutine's accuracy by comparing its results to the known property of hydrogen atom as shown below.

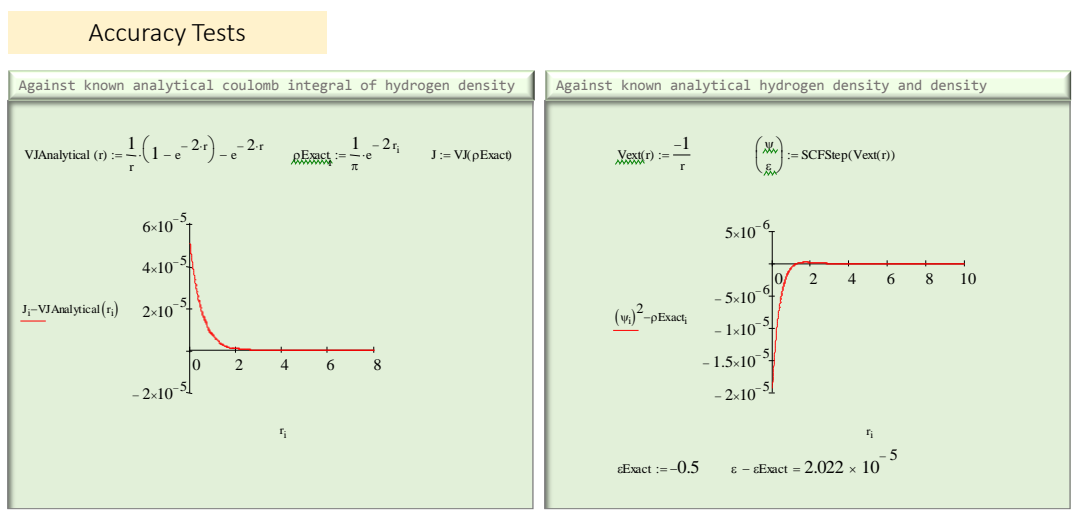

On the left was the test for the Coulomb integral, which was needed to compute Eq. (10) of their paper. We used trapezoid integration method which gave only 0.05 mHartree error compared to the known analytical value for hydrogen atom.

One the right was the test for electron density. If the hydrogen Coulomb potential $v(r) = -\frac{1}{r}$ was fed into our subroutine, it calculated the wave functional $\psi(r)$ whose squared absolute could be readily compared to the known electron density of hydrogen. The error was only 2x10$^{-5}$ au.

Assuming the errors discussed above were of the same order of magnitude when applying the code to the Hooke atom, it was sufficiently accurate to draw conclusive statement from the figure 1.

Specifically, we transformed the Eq. (9) in their paper into a matrix form using finite difference method. Using spherical symmetry and the ground state's zero angular momentum, the 3-dimensional partial differential equation,

$$\left[-\frac{1}{2}\nabla^2 + v(r)\right]\psi(r) = \epsilon\psi(r) \tag{S18}$$

was rewritten into the 1-dimentional (radial) finite difference form

$$-\frac{1}{2}(1 - \frac{h}{r_i})\psi_{i-1} + (1 + h^2 v_i)\psi_i - \frac{1}{2}(1 + \frac{h}{r_i})\psi_{i+1} = (\epsilon h^2)\psi_i, \tag{S19}$$

where "h" is the step size of a uniform radial grid points. Explicitly writing the Eq. (S19) for a few grid points, one began to see a useful pattern.

```
i=0 @ r = h     0·ψ₋₁ + (1 + h²v₀)ψ₀ − 1·ψ₁ + 0·ψ₂ + 0·ψ₃ + 0·ψ₄ + ⋯      = (εh²)ψ₀
i=1 @ r = 2h    −1/4·ψ₀ + (1 + h²v₁)ψ₁ − 3/4·ψ₂ + 0·ψ₃ + 0·ψ₄ + ⋯         = (εh²)ψ₁
i=2 @ r = 3h    0·ψ₀ − 1/3·ψ₁ + (1 + h²v₂)ψ₂ − 2/3·ψ₃ + 0·ψ₄ + ⋯          = (εh²)ψ₂
```

$$\tag{S20}$$

On the first line for r=h, note how the coefficient of $\psi_{-1}$ was zero. With our index notation, i=0 meant r=h, then i=(-1) was at the origin r=0. The fact that $\psi_{-1}$'s coefficient was zero meant that the wave function at the origin was decoupled (only for this numerical scheme) from the rest of the grid points.

Therefore, for this method, we did not need to include $\psi$ at the origin into the calculation. Our grid points were $r_i \in \{h, 2h, 3h, \cdots\}$. Not having to include the origin into the calculation had an obvious advantage because it allowed us to numerically solve for the wave function even though the potential was singular at the origin.

The set of equations listed in Eq. (S20) could be casted into a matrix form,

$$\begin{pmatrix} 1+h^2 v_0 & -1 & & & 0 \\ -1/4 & 1+h^2 v_1 & -3/4 & & \\ & -1/3 & 1+h^2 v_2 & -2/3 & \\ & & & \ddots & \\ 0 & & & & \end{pmatrix} \begin{pmatrix} \psi_0 \\ \psi_1 \\ \psi_2 \\ \vdots \end{pmatrix} = \epsilon' \begin{pmatrix} \psi_0 \\ \psi_1 \\ \psi_2 \\ \vdots \end{pmatrix},$$

or

$$\widetilde{M}\vec{\psi} = \epsilon'\vec{\psi}. \qquad (S21)$$

One could then use available software package such as Matlab, Octave, or MathCAD to solve for the eigen value $\epsilon'$ and the eigen vector $\vec{\psi}$. We implemented the method into MathCAD software package as shown below.

$$VJ(\rho) := \begin{vmatrix} \text{for } n \in 0..(N-1) \\ V_n \leftarrow 4\cdot\pi\cdot\frac{h}{2}\cdot\left[\frac{1}{r_n}\cdot\left[\sum_{i=0}^{n}\left[2\cdot(r_i)^2\cdot\rho_i\right] - (r_n)^2\cdot\rho_n\right] + \sum_{i=n}^{N-1}(2\cdot r_i\cdot\rho_i) - r_n\cdot\rho_n - r_{N-1}\cdot\rho_{N-1}\right] \\ \text{return } V \end{vmatrix}$$

$$SCFStep(Veff) := \begin{vmatrix} \text{for } i \in 0..(N-1) \\ \quad \text{for } j \in 0..(N-1) \\ \quad\quad M_{i,j} \leftarrow 0 \\ \quad\quad M_{i,j} \leftarrow 1 + h^2 \cdot Veff_i \text{ if } i = j \\ \quad\quad M_{i,j} \leftarrow -\left[\frac{1}{2}\cdot\left(1 - \frac{h}{r_i}\right)\right] \text{ if } 1 = i - j \\ \quad\quad M_{i,j} \leftarrow -\left[\frac{1}{2}\cdot\left(1 + \frac{h}{r_i}\right)\right] \text{ if } 1 = j - i \\ \text{eigenval} \leftarrow \text{sort}(\text{eigenvals}(M)) \\ \psi \leftarrow \text{eigenvec}(M, \text{eigenval}_0) \\ \text{Norm} \leftarrow \dfrac{1}{\sqrt{4\pi\cdot h\cdot\sum_{i=0}^{N-1}\left[(r_i)^2\cdot(\psi_i)^2\right]}} \\ \psi \leftarrow \text{Norm}\cdot\psi \\ \text{return } \begin{pmatrix} \psi \\ \dfrac{\text{eigenval}_0}{h^2} \end{pmatrix} \end{vmatrix}$$

$$SCFConv(maxCycle, \psi) := \begin{vmatrix} \varepsilon Now \leftarrow 0 \\ \text{for } i \in 0..maxCycle \\ \quad \begin{pmatrix} \psi \\ \varepsilon \end{pmatrix} \leftarrow SCFStep\left(Vext(r) + \frac{1}{2}\cdot VJ(2\psi^2)\right) \\ \quad \text{break if } |\varepsilon - \varepsilon Now| < 10^{-6} \\ \quad \varepsilon Now \leftarrow \varepsilon \\ \text{return } \begin{pmatrix} \psi \\ \varepsilon \\ i \end{pmatrix} \end{vmatrix}$$

**Set Parameters**

$R := 10$  $\quad Vext(r) := \frac{1}{2}\cdot\frac{1}{4}\cdot r^2$  $\quad N := 800$

$h := \dfrac{R}{N+1}$  $\quad i := 0..(N-1)$  $\quad r_i := h\cdot(i+1)$  $\quad \psi_i := 0$

**Main Commands**

$\begin{pmatrix} \psi \\ \varepsilon \\ nCycle \end{pmatrix} := SCFConv(100, \psi)$  $\quad nCycle = 8$  $\quad \rho HF_i := 2\cdot(\psi_i)^2$

The function "SCFStep(Veff)" took the specified potential and built the matrix $\widetilde{M}$. It then solved for the lowest eigen value, found the corresponding eigen function, and returned them as outputs.

The function "SCFConv" ran the calculation iteratively until the eigen energy no longer changed (0.001 mHartree threshold). Note how it took the wave function from the previous cycle, computed the Coulomb potential, and fed it back into the next iteration.

In this work, we used 800 grid points, which took 8 cycles to complete. We then used the Hartree-Fock wave function to compute the electron density and the graphs in figure 1.

S21# References

1. Chachiyo, T. *et al.* "Siam-Quantum: a compact open-source quantum simulation software for molecules," Thailand. https://sites.google.com/site/siamquantum (2016).

2. Chachiyo, T. Communication: Simple and accurate uniform electron gas correlation energy for the full range of densities. *J. Chem. Phys.* **145,** (2016).

3. Wang, Y. & Perdew, J. P. Spin scaling of the electron-gas correlation energy in the high-density limit. *Phys. Rev. B* **43,** 8911–8916 (1991).

4. Barbieri, P. L., Fantin, P. A. & Jorge, F. E. Gaussian basis sets of triple and quadruple zeta valence quality for correlated wave functions. *Mol. Phys.* **104,** 2945–2954 (2006).

5. Schuchardt, K. L. *et al.* Basis set exchange: A community database for computational sciences. *J. Chem. Inf. Model.* **47,** 1045–1052 (2007).

6. Schmidt, M. W. *et al.* General atomic and molecular electronic structure system. *J. Comput. Chem.* **14,** 1347–1363 (1993).

7. Curtiss, L. A., Raghavachari, K., Trucks, G. W. & Pople, J. A. Gaussian-2 theory for molecular energies of first- and second-row compounds. *J. Chem. Phys.* **94,** 7221–7230 (1991).

8. Kohn, W. & Sham, L. J. Self-Consistent equations including exchange and correlation effects. *Phys. Rev.* **140,** A1133 (1965).

9. O'neill, D. P. & Gill, P. M. W. Benchmark correlation energies for small molecules. *Mol. Phys.* **103,** 763–766 (2005)

10. Kais, S., Herschbach, D. R., Handy, N. C., Murray, C. W. & Laming, G. J. Density functionals and dimensional renormalization for an exactly solvable model. *J. Chem. Phys.* **99**, 417–425 (1993).